\documentstyle[12pt]{article}
\begin{document}
\begin{titlepage}

\center{{\bf SIMILARITY SOLUTIONS AND COLLAPSE IN THE ATTRACTIVE
GROSS-PITAEVSKII EQUATION}} \vspace{1cm} {\center{\bf A.V.
Rybin${}^{\dag}$, G.G. Varzugin${}^{\ast}$, M.
Lindberg${}^{\ast\ast}$ \\ J. Timonen${}^{\dag}$
 and
R.K. Bullough${}^{ \ddag}$}}

\begin{center}{\dag Department of Physics, University of Jyv{\"a}skyl{\"a}}\\
 {PO Box 35, FIN-40351}\\
{Jyv{\"a}skyl{\"a}, Finland}
\end{center}

\begin{center}{$\ast$ Institute of Physics}\\
 {St. Petersburg State University}\\
 {198904, St. Petersburg, Russia}
\end{center}

\begin{center}{${}^{\ast\ast}$  Department of Physics, \AA bo Akademi University\\
 Porthansgatan 3, FIN-20500\\
\AA bo, Finland}
\end{center}

 \begin{center}
{\ddag Department of Mathematics}, {UMIST}\\
 {PO Box 88}
{Manchester M60 1QD}, {UK}
\end{center}

 \abstract{We analyse a generalised Gross-Pitaevskii equation involving a
paraboloidal trap potential in $D$ space dimensions and
generalised to a nonlinearity of order $2n+1$. For {\em
attractive} coupling constants collapse of the particle density
occurs for $Dn\ge 2$ and typically to a $\delta$-function centered
at the origin of the trap. By introducing a new dynamical variable
for the spherically symmetric solutions we show that all such
solutions are self-similar close to the center of the trap.  {\em
Exact} self-similar solutions occur if, and only if, $Dn=2$, and
for this case of $Dn=2$ we exhibit an exact but rather special
$D=1$ analytical self-similar solution collapsing to a
$\delta$-function which however recovers and collapses
periodically, while the ordinary G-P equation in 2 space
dimensions also has a special solution with periodic
$\delta$-function collapses and revivals of the density. The
relevance of these various results to attractive Bose-Einstein
condensation in spherically symmetric traps is discussed.}


 communicating author: Andrei.Rybin@phys.jyu.fi
\end{titlepage}

The experimental discovery of Bose-Einstein condensation (BEC) in
trapped vapours of cooled alkali atoms~\cite{be1,be2,be3,be4} has
opened up unique possibilities for the investigation of collective
many-body effects in dilute gases. In the experiments the cloud of
atoms is isolated from the environment by a  magnetic trap. After
cooling the cloud exhibits Bose-Einstein condensation i.e. the
existence of a macroscopically populated quantum state. The study
of the dynamics of this quantum state is an important fundamental
problem in many-body quantum physics. For three space dimensions
$D=3$ the dynamics of the condensate can be described within the
Hartree-Fock approximation by the Gross-Pitaevskii equation

\begin{equation}
i \hbar\Phi_t +
\frac{\hbar^2}{2m}\Delta_x\Phi-\frac{4\pi\hbar^2a_s}{m}\Phi|\Phi|^2-
V(\vec{x})\Phi=0,\label{GP}
\end{equation}

 where $\Phi(\vec{x},t)$ is the wave function of the condensate, the external potential $V(\vec{x})$ models the wall-less
confinement (the trap), $m$ is the mass of an individual atom,
$a_s$ is the scattering length, and
$\Delta_x=\sum_i^3\frac{\partial^2}{\partial x_i^2}$ is the
Laplace operator. A convenient choice for the confining trap is
the paraboloidal potential assumed here to be spherically
symmetric for simplicity, i.e. $V=\frac{m
\omega_0^2}{2}\vec{x}^2$.

In this paper we are concerned with condensates in  $D=3$ and
$D=2$ dimensions. The Bose-Einstein condensate in two space
dimensions is
 only marginally stable in that below the critical temperature correlations decay,
 but decay  only as a power law~\cite{kad,bog}. Recent experimental techniques
allow realisation of a two-dimensional trap for e.g.
spin-polarized hydrogen adsorbed on a helium
surface~\cite{ex1,ex2}. The  dynamics of  trapped Bose-Einstein
condensates and the search for the related soliton-like solutions
of the Gross-Pitaevskii equations is thus an interesting and
relevant problem also in two dimensions. In this paper we
concentrate on some aspects of this dynamics and on the existence
of self-similar solutions of the Gross-Pitaevskii equation in
particular. Self-similarity is an important and useful concept in
nonlinear dynamics, particularly so when collapsing systems are
being  considered ~\cite{zakh,wad2} as they are below. This
phenomenon of collapse appears in Bose-Einstein condensates with
negative scattering length, as for example in ${}^7Li$ (see
e.g.~\cite{pit}).   In this paper we show that self-similar
behaviour   only appears in two-dimensional traps although
'attractive' condensates ($a_s<0$) collapse for all $D\ge 2$.

To begin with we consider a generalized $D$-dimensional
Gross-Pitaevskii equation, which for units such that $\hbar=1$,
$m=1/2$ can be expressed in the form

\begin{equation}
i \psi_t +\Delta_x\psi-2\kappa\psi |\psi|^{2n}-\frac{\omega^2}{4}
r^2\psi=0.\label{GP1_a}
\end{equation}
Here $\Delta_x$ is the $D$-dimensional Laplace operator and
$r^2=\sum_i^D x_i^2$. Notice that "generalisation" means here an
exponent $2n$ instead of the $2$ which appears in the ordinary G-P
equation.
%For the most realistic case $D=3$, $n=1$ the variables
%in Eq.~(\ref{GP1_a}) can be introduced as follows
%\begin{equation}
%x_i=\sqrt{\frac{2}{\omega}}\frac{y_i}{a_{\scriptscriptstyle{HO}}},\,
%t=\frac{\omega_0}{\omega}\tau,\,\psi=a_{\scriptscriptstyle{HO}}^{\frac{3}{2}
%}\Phi,\, \kappa=2\pi\omega
%\frac{a_s}{a_{\scriptscriptstyle{HO}}}.\label{phys}
%\end{equation}

%Here $a_{\scriptscriptstyle{HO}}=\sqrt{\frac{\hbar}{m\omega_0}}$
%is the characteristic length of the harmonic oscillator and
%$\omega$ is a dimensionless parameter.

%INSERTION_BEGINNING

We consider only the attractive case of Eq.~(\ref{GP1_a})
$\kappa<0$ and the boundary conditions are vanishing at infinity.
An  observation is that a   symmetry which leaves
Eq.~(\ref{GP1_a}) invariant is

\begin{equation}
\psi(\vec{x},t)\to  e^{i\left\{ \frac{\omega}{4}\sin(\omega
t+\varphi_0)(2\vec{x}\cdot\vec{\eta_0}+\vec{\eta_0}\cdot\vec{\eta_0}\cos(\omega
t+\varphi_0))\right\}}\psi(\vec{x}+\vec{\eta_0}\cos(\omega
t+\varphi_0),t). \label{symmetry}
\end{equation}

in which $\vec{\eta_0}$ is an arbitrary vector in $D$ dimensions,
$\varphi_0$ is an arbitrary phase. This symmetry  reveals the, in
general,  {\em oscillatory} character of the wave packet dynamics
of Eq.~(\ref{GP1_a}) whether $\kappa>0$ or $\kappa<0$. In
Ref.~\cite{zakh} and its references  'collapse' was demonstrated
for $\omega=0$ and $\kappa<0$. Solutions  become singular in a
final time interval if the condition
\begin{equation}
nD\ge 2 \nonumber
\end{equation}
is fulfilled. We show here how the same condition arises in the
present context, where $\omega\neq 0$ (and $\kappa<0$).
Ref.~\cite{wad} has addressed the same problem (of $\omega\neq0$)
for $n=1$ and $D=2$ and $D=3$. Following both~\cite{zakh} and
\cite{wad} we use the functional $U[\psi]=\int_{R^D}r^2|\psi|^2
d^Dx$ in which $r=|\vec{x}|:~U[\psi]\ge 0$. From Eq.~(\ref{GP1_a})
this functional satisfies a second order ordinary differential
equation whose solution is
\begin{eqnarray}
\label{potential} &&U[\psi]={4\sin^2(\omega
t)\over\omega^2}E_{NLS}+U_0\cos^2(\omega t) +J_0{\sin(2\omega
t)\over 2\omega}\nonumber\\ &&+{4\kappa (Dn-2)\over\omega(n+1)}
\int\limits_0^t\sin(2\omega(t-t^\prime))I_{2n+2}[\psi]dt^\prime
\label{main_1}
\end{eqnarray}
with

\begin{equation}U_0=U[\psi]|_{t=0},\quad J_0={d\over dt}U[\psi]|_{t=0},\quad
E_{NLS}=E[\psi]-{\omega^2\over 4}U_0,\label{ENLS}\end{equation}
$$I_q[\psi]=\int_{R^D}|\psi|^q d^Dx,$$ where
\begin{equation}E[\psi]=\int_{R^D}\left(|\nabla\psi|^2+ {2\kappa \over
n+1}|\psi|^{2n+2} +{\omega^2\over 4}r^2|\psi|^2
\right)d^Dx\label{energia}\end{equation}  is an obvious "energy"
functional and is the Hamiltonian of Eq.~(\ref{GP1_a}) with the
bracket
$\{\psi(\vec{x}),\psi^\ast(\vec{y})\}=i\delta(\vec{x}-\vec{y})$.

Hamiltonian Eq.~(\ref{energia}) is a constant of the motion fixed
by the initial data. For  $\kappa<0$ and smooth enough initial
data it is not bounded below  while $E_{NLS}$ as defined in
Eqs.~(\ref{ENLS})  has the same properties. The condition
$E_{NLS}\le 0$, for example, still admits a large amount of
physically accessible initial data. A second constant of the
motion is $\int_{R^D}|\psi|^2d^Dx\equiv {\cal N}$, the total
number of bosons (atoms). Careful scrutiny of $U[\psi]$ of
Eq.~(\ref{main_1}) then shows (see also \cite{zakh}, \cite{wad})
that provided that
\begin{equation}\kappa<0,\,\, Dn\ge 2,\,\,
E_{NLS}\le 0,\label{col}\end{equation} with the exception of the
special case $Dn=2$, $E_{NLS}=J_0=0$, there is always at least one
point $t=t_\ast\in\left(0,\frac{\pi}{2\omega}\right]$ such that
the right hand side  of Eq.~(\ref{main_1}) becomes negative for
$t>t_\ast$. Since by its definition the functional $U[\psi]$ is
nonnegative, this contradiction leads to the conclusion that
$\psi$ cannot be continued beyond the point $t=t_\ast$ and must
exhibit a singularity. We show below that this singularity is
typically $|\psi|^2\to {\cal N}\delta(\vec{x})$. However, for the
special case $Dn=2$, $E_{NLS}=J_0=0$, the functional
$U[\psi]=U_0\cos^2(\omega t)$  never becomes negative. We show
below that collapse in $|\psi|^2$ occurs with $|\psi|^2 \to {\cal
N}\delta(\vec{x})$ as $t \to t_\ast$, but now this can be followed
by revival and periodic collapse of period $\pi/\omega$. There is
some evidence that a form of collapse could occur in general even
when $U[\psi]$ apparently remains positive, i.e. at some point
$t<t_\ast$ (see \cite{rasmussen,rypdal} and references therein
where $\omega\equiv 0$). We shall assume here that collapse occurs
only at a zero of $U[\psi]$.

Thus the conditions Eq.~(\ref{col}) are sufficient for $U[\psi]$
to reach a zero at $t=t_\ast\le \frac{\pi}{2\omega}$ and,
generically at least, $|\psi|^2\to {\cal N}\delta(\vec{x})$ there.
These conditions are sufficient but not necessary: for given such
evolution for $E_{NLS}\le 0$, the transformation
Eq.~(\ref{symmetry}) can increase $E_{NLS}$ to $>0$ while the
evolution remains singular. This is true for example for the exact
analytical solution Eq.~(\ref{sol1}) for $Dn=2$ we give below. The
formation of these singularities may be very sensitive to the
initial conditions and the values of the parameters. Evidently
these results mean that for $E_{NLS}\le 0$ initially collapse and
blow-up will occur for all ${\cal N}\ge {\cal N}_c$~\cite{wad}
(see also~\cite{lem} and references therein).
%For $D=3$ and Gaussian initial data in particular, one finds the
%actual number ${\cal
%N}_c=\frac{3}{4}\sqrt{2\pi}\frac{a_{\scriptscriptstyle{HO}}}{|a_s|}$,
%where $a_{\scriptscriptstyle{HO}}$ is the quantum oscillator
%length
%$a_{\scriptscriptstyle{HO}}=\sqrt{\frac{\hbar}{m\omega_0}}$.  For
%${}^7Li$ as example ${\cal N}_c=2800$, which compares with the
%result found in.

We turn to the problem of similarity solutions which within the
terms of our analysis arise only for $Dn=2$. We seek spherically
symmetric solutions of Eq.~(\ref{GP1_a}) in the form
$\psi(r,t)=A(r,t)e^{i\phi(r,t)}$ in which $r=|\vec{x}|$. From
Eq.~(\ref{GP1_a}) we arrive at the set of equations
\begin{equation}
\frac{\partial A^2}{\partial t}+
\frac{2}{r^{D-1}}\frac{\partial}{\partial r}\left(r^{D-1}A^2\frac{\partial\phi}{\partial r}
\right)=0
\label{RGP1}
\end{equation}
\begin{equation}
\frac{1}{r^{D-1}}\frac{\partial}{\partial
r}\left(r^{D-1}\frac{\partial A} {\partial
r}\right)-\left(\frac{\partial\phi}{\partial t}
+\left(\frac{\partial\phi}{\partial
r}\right)^2+\frac{\omega^2}{4}r^2\right)A- 2\kappa A^{2n+1}=0.
\label{RGP2}
\end{equation}

It is not evident how similarity solutions could be constructed
from this set of equations in the general case, and we therefore
choose to make an ansatz for the amplitude variable $A(r,t)$:
\begin{equation}
A(r,t)=\left(\frac{\eta(r,t)}{r}\right)^{D-1\over 2}
\left(\frac{\partial\eta(r,t)}{\partial
r}\right)^{1\over2}A_0(\eta(r,t)). \label{ansatz1}
\end{equation}

This ansatz solves Eq.~(\ref{RGP1}) provided that the function
$\eta$ satisfies

\begin{equation}
\frac{\partial\eta}{\partial t}+2\frac{\partial\eta}{\partial r}
\frac{\partial\phi}{\partial r}=0. \label{oneta2}
\end{equation}
Notice that the function $A_0(\eta)$ is arbitrary and the ansatz~
Eq.(\ref{ansatz1}) describes an {\em arbitrary} spherically
symmetric solution. The gradient $\frac{\partial\phi}{\partial r}$
is related to the velocity of the particles of the condensate,
and, through Eq.~(\ref{oneta2}), $\eta(r,t)$ is then related  to
the local time dependent concentration of condensate particles. In
fact $\eta(r,t)$ completely determines this concentration as is
evident from the number of particles $n(r,t)$ in the interval
$[0,r]$, which is
\begin{equation}
n(r,t)\equiv
\Omega_D\int\limits_0^rr^{D-1}A^2(r,t)dr=\Omega_D\int\limits_0^\eta
\eta^{D-1} A^2_0(\eta
)d\eta,\,\,\Omega_D=\frac{2\pi^{D/2}}{\Gamma(D/2)},
\end{equation}
while $n(\infty,t)={\cal N}$ is independent of $t$.
 From the ansatz Eq.~(\ref{ansatz1}) we can deduce that $\eta(r,t)$
 is a
monotonically increasing function of  $r$ i.e. $\frac{\partial
\eta}{\partial r}>0$, and $\eta\rightarrow\infty$ when
$r\rightarrow\infty$. Also, in the vicinity of the origin $r=0$,
  $\eta$ behaves as
\begin{equation}
\eta(r,t)=r/\rho(t)+O(r^2), \label{boundary1}
\end{equation}
where $\rho(t)$ is a function of time.
 The solution for $\eta(r,t)$ is self-similar if
$\eta=r/\rho(t)$ exactly. This 'self-similarity' is in the sense
that the function $\eta$ depends now on a single variable
$\eta=r/\rho(t)$. From Eq. (\ref{oneta2}) it follows immediately
that in this case the {\em phase} $\phi(r,t)$ is
 quadratic in $r$,
\begin{equation}
\phi(r,t)=\phi_0(t)+{1\over
4}\frac{\rho^\prime(t)}{\rho(t)}r^2.\label{faza}
\end{equation}

The Eq.~(\ref{RGP2}) should now be understood  as an equation for
$A_0$. Consider first the case $nD=2$. Separating the variables in
this equation we find that
\begin{equation}
\frac{1}{\eta^{D-1}}\frac{\partial}{\partial\eta}\left(\eta^{D-1}
\frac{\partial A_0}{\partial\eta}\right)-2\kappa A_0^{2n+1}-
(\mu+\lambda\eta^2)A_0=0
\label{selfsim1}
\end{equation}
\begin{equation}
\phi_0^\prime+\frac{\mu}{\rho^2}=0
\end{equation}
\begin{equation}
\rho^{\prime\prime}+\omega^2\rho-\frac{4\lambda}{\rho^3}=0.
\label{selfsim3}
\end{equation}
Here $\lambda$ and $\mu$ are arbitrary constants.

A solution of Eq.~ (\ref{selfsim3}) can easily be found in the
form
\begin{equation}
\rho(t)=\sqrt{\cos^2(\omega
t)+\frac{4\lambda}{\omega^2}\sin^2(\omega t)}.\label{sol_rho}
\end{equation}
Other solutions can be obtained through the  transformation $t\rightarrow t+t_0$ and
$\rho(t)\rightarrow h(t)\rho(s(t))$, where

\begin{equation}h(t)=(\sqrt{1+\alpha^2}+\alpha\cos(2\omega
t))^{1\over2},\, s(t)=
\frac{1}{\omega}\tan^{-1}\left((\sqrt{1+\alpha^2}-\alpha)\tan(\omega
t)\right).\label{xx}\end{equation} We have thus demonstrated that
for $nD=2$, $\eta(r,t)=r/\rho(t)$ with $\rho(t)$ given by
Eqs.~(\ref{sol_rho}) and (\ref{xx}) is indeed a solution, and
Eqs.~(\ref{ansatz1}) and (\ref{faza}) now provide the
corresponding self-similar solution of the Gross-Pitaevskii
equation Eq.~(\ref{GP1_a}). For the explicit form of this solution
one still needs to solve Eq.~(\ref{selfsim1}) for $A_0(\eta)$.

 In the case  $nD\neq2$ there
are no self-similar solutions (except the trivial case
$\rho=$const). Indeed, for the existence of such solutions we need
to require that both $A_0^{2n}$ and $\Delta_\eta A_0/A_0$ are
functions quadratic in $\eta$. These conditions  cannot obviously
be satisfied. This means that even though the solution given by
Eq.~(\ref{ansatz1}) is locally self-similar for any $D$ in the
vicinity of   $r=0$, the exact self-similarity  is only realised
  for $Dn=2$.

For the self-similar solutions there are two integral identities.
Multiplying Eq.~(\ref{selfsim1}) by $\eta^{D-1}A_0$ and by
$\eta^D\partial A_0/\partial\eta$, respectively, and integrating
by parts, we find after a little algebra that
\begin{equation}
\int\limits_0^\infty d\eta\eta^{D-1}\left( \left(\frac{\partial
A_0}{\partial\eta}\right)^2+
\frac{2\kappa}{n+1}A_0^{2n+2}-\lambda\eta^2A_0^2\right)=0
\label{int1}
\end{equation}
\begin{equation}
\int\limits_0^\infty d\eta \eta^{D-1}\left( \left(\frac{\partial
A_0}{\partial\eta}\right)^2+
\kappa\frac{n+2}{n+1}A_0^{2n+2}+\frac{1}{2}\mu A_0^2\right)=0.
\end{equation}
Using the identity  Eq.~(\ref{int1}) we easily find that the total
energy of the solution, $E[\psi]$, Eq.~(\ref{energia})
 is given by
\begin{equation}
E[\psi]=\frac{1}{4}e(\rho)\int\limits_0^\infty\eta^{D+1}A_0^2(\eta)d\eta,
\quad
e(\rho)=(\rho^\prime)^2+\omega^2\rho^2+\frac{4\lambda}{\rho^2}.\label{int3}
\end{equation}

As an  example of an exact solution of Eq.~(\ref{selfsim1}) we
consider here the attractive generalised G-P equation in one
dimension: $D=1, n=2, \lambda=0$ and $\kappa<0$. In this case we
find that
\begin{equation}
A_0(\eta)=\frac{p_0} {\sqrt{\cosh\left(\frac{2}{3}\sqrt{6|\kappa
|}p_0^2\eta\right)}}, \quad \mu=\frac{2}{3}|\kappa|p_0^4,
\end{equation}
and the solution of Eq.~(\ref{GP1_a}) can be expressed in the form
\begin{equation}
\psi(x,t) =\frac{p_0}{\sqrt{\cos(\omega
t)}}\frac{\exp{\left\lbrace -i\tan{(\omega
t)}\left(\frac{\omega}{4} x^2 - \frac{2}{3\omega}|\kappa
|p_0^4\right)\right\rbrace}}{\sqrt{\cosh{
\left(\frac{2}{3}\sqrt{6|\kappa |}p_0^2\frac{x}{\cos{(\omega
t)}}\right)}}}.\label{sol1} \label{sol}
\end{equation}
For an attractive condensate ($\kappa<0$) we expect the solution
to become singular at a finite time.
 But it is indeed obvious that the solution Eq.~(\ref{sol1})
becomes singular for $t\to \frac{\pi}{2\omega}$ when its amplitude
diverges as $1/\sqrt{\frac{\pi}{2\omega}-t}$.   In this limit
$|\psi|^2\to
\frac{\pi}{2}{\sqrt{\frac{3}{2|\kappa|}}}\delta(x)={\cal
N}\delta(x)$ which is the convergence to the $\delta$-function
expected.
%This
%behavior is only possible if $\lambda\le 0$ (cf.~Eq.(\ref{int1})
%with the requirement $E_{NLS}\le0$ and $\kappa<0$ ).
Notice that $|\psi|^2$ from Eq.~(\ref{sol}) is now periodic of
period $\frac{\pi}{\omega}$ while the solution Eq.~(\ref{sol})
itself has the jumps in  phase, compounded by branch point
singularities, when crossing the singularities of $|\psi|$ at
$t=\frac{\pi}{2\omega}(2k+1)$, $k\in {\bf Z}$\cite{foot}.

 If  now we simply
assume that the point of collapse is $t=t_\ast$ defined below
Eq.~(\ref{col}), it can still be shown that the collapse occurs to
a $\delta$-function centered on the trap. A consideration leading
to this conclusion is the following: the equality $U[\psi]=0$
means that $|\psi(\vec{x},t_\ast)|=0$ for any $\vec{x}$ except
possibly at the origin. Since ${\cal N}=\int d^D x|\psi|^2$ is a
constant of motion   identified as the total number of bosons
(atoms), the obvious physical solution is $|\psi|^2={\cal
N}\delta(\vec{x})$ excluding other possible generalised functions.
 The spherically symmetric case can be treated rigorously.
 Consider for this the functional $U[\psi]$ taken on the
ansatz Eq.(\ref{ansatz1}), i.e.
\begin{equation}U[\psi]=\int\limits_0^\infty d\eta
r^2(\eta,t)\eta^{D-1} A_0^2(\eta).\label{u_an}\end{equation} For
$U[\psi]=0$ it immediately follows from Eq.~(\ref{u_an})  that
$r(\eta,t_\ast)=0$. For an appropriate arbitrary test  function
$\varphi(r)$ consider now the limit \begin{eqnarray}&& \lim_{t\to
t_\ast}\Omega_D\int^\infty_0drr^{D-1}|\psi(r,t)|^2\varphi(r)\nonumber\\&&=
\lim_{t\to
t_\ast}\Omega_D\int^\infty_0d\eta\eta^{D-1}|A_0(\eta)|^2\varphi(r(\eta,t))={\cal
N}\varphi(0). \end{eqnarray} This result  means rigorously  that
for spherically symmetric solutions for which $U[\psi]$ evolves to
a zero at $t=t_\ast$.
  the system 'blows-up' to the
$\delta$-function singularity
 $$\lim_{t\to t_\ast}|\psi(r,t)|^2={\cal N}\delta(r).$$
This result is  particularly evident for the self-similar
solutions for which $\eta=r/\rho$. In this case

$$U[\psi]=\left(\cos^2(\omega
t)+\frac{4\lambda}{\omega^2}\sin^2(\omega t)\right)\int_0^\infty
d\eta \eta^{D+1}A_0^2(\eta) $$ and the point of collapse for
$\lambda\le 0$ ($E_{NLS}\le 0$) can be readily found as
$t_\ast=({1}/{\omega})\tan^{-1}\left({\omega}/{2\sqrt{|\lambda|}}\right)$.

Notice again that when $\lambda=0$ the functional $U$ never
becomes negative and there is a  possibility of periodic
$\delta$-function collapses and revivals of the condensate density
in this case of two-dimensional traps.

Even though the methods are different, some part of the results
reported here
 is   analogous to that obtained in
Refs.~\cite{zakh, rasmussen, rypdal} for the Nonlinear
Schr\"odinger equation (NLS), which is the Gross-Pitaevskii
equation with $ \omega\equiv 0$. This analogy is related to the
fact that for $Dn=2$ the generalised NLS and GP equations are
equivalent. For the change of variables~\cite{ryb}
\begin{equation}
\theta=\frac{1}{\omega}\tan(\omega t),\,z_i=\frac{x_i}{\cos(\omega t)}
\label{ch1D}
\end{equation}
\begin{equation}
\psi( x, t)=(\cos(\omega t))^{-\frac{D}{2}}
\exp\left\{-i\frac{\omega}{4}\tan(\omega t)r^2\right\}
p(z,\theta)
\label{ch2D}
\end{equation}
maps Eq.(\ref{GP1_a}) to
\begin{equation}
i p_\theta +\Delta_z p-\frac{2\kappa}{(1+\omega^2\theta^2)^{-\frac{Dn}{2}+1}}
 p| p|^{2n} =0,\label{GPT}
\end{equation}
and it is clear that for $Dn=2$ the $\theta$ dependence of the
effective 'coupling constant' disappears and the NLS system is
recovered. This means in particular that the whole variety of
results available for the two dimensional NLS for $n=1$ is
directly applicable to the Gross-Pitaevskii equation for $n=1$ in
two space dimensions. It is interesting that the Gross-Pitaevskii
equation only allows a self-similar solution of the type
considered in this paper in this case when it can be exactly
transformed to the NLS equation.

It is worth mentioning that for $Dn=2$ all self-similar solutions
of the Gross-Pitaevskii equation are invariant under the
transformation
\begin{equation}
\psi(x,t)\rightarrow h(t)^{-{D\over2}}\exp
\left(\frac{ih^\prime(t)}{4h(t)}r^2\right) \psi\left({x\over
h(t)}, s(t) \right). \label{sym2}
\end{equation}
For the solution Eq.~(\ref{sol}) this transformation means a mere
rescaling  $p_0\rightarrow
p_0/(\alpha+\sqrt{1+\alpha^2})^{1\over4}$.

We emphasize  that our similarity analysis of the Gross-Pitaevskii
equation is based on the ansatz Eq.~(\ref{ansatz1}). This approach
is applicable to the Gross-Pitaevskii equation in $D$ space
dimensions and with an arbitrary external potential $V(\vec{x})$.
It can also be shown that the dynamics described by the
Gross-Pitaevskii equation for an arbitrary initial condition which
has an extremum, is effectively equivalent to a system describing
a $D$-dimensional classical particle.  This dynamical system
generalises that found in ~\cite{gar} for Gaussian initial
profiles through a variational approach. These results will be
reported in a forthcoming publication.

We showed in this paper that for $Dn=2$ alone the generalised
Gross-Pitaevskii Eq.~(\ref{GP1_a}) equation allows self-similar
solutions, and that in this case it can be exactly transformed to
the NLS equation with no trap potential. An explicit solution was
given for $D=1$, $n=2$, $\kappa<0$ which displayed a delta
function divergence at a finite time. We further showed, for
$Dn\ge 2$ and $\kappa<0$, that {\em all} spherically symmetric
solutions with $E_{NLS}\le 0$ collapse in a finite time to a
$\delta$-function centered at the origin of the trap while we
showed generally that even without such symmetry evolution may be
to the $\delta$-function singularity. The ordinary
Gross-Pitaevskii equation in $2$ space dimensions and with
$\kappa<0$, $E_{NLS}=0$, was shown to have periodic
$\delta$-function collapses and subsequent {\em revivals} of the
particle density.

One of us (AVR) wishes  to thank M. Wadati, J. Hietarinta and S.
Jaakkola with collaborators for useful discussions.

GGV was partly supported by a Russian Federation research grant
RFBR No 98-01-01063


\begin{thebibliography}{99}
\bibitem{foot} Physically Eq.~(\ref{sol1}) must be a  special
solution related to the particular choice $\lambda=0$. The
normalisation of this solution is independent of the parameters
and indeed one checks that ${\cal N}$ must take the particular
value ${\cal N}=\frac{\pi}{2}{\sqrt{\frac{3}{2|\kappa|}}}$.
\end{thebibliography}
\end{document}